\documentclass[twocolumn,preprint]{jpsj2}
\usepackage{graphicx}

\def\runtitle{A which--way device}
\def\runauthor{Angus \textsc{MacKinnon} and Andrew D. \textsc{Armour}}

\title{%
Quantum Interference and Inelastic Scattering in a Model Which--Way Device
}

\author{%
Angus \textsc{MacKinnon}$^{1,2,}$\thanks{Present address: Blackett Lab.,
Imperial College London, London SW7 2BW, UK; E-mail: a.mackinnon@imperial.ac.uk} and
Andrew D. \textsc{Armour}$^{3}$
}

\inst{%
$^{1}$ The Blackett Laboratory, Imperial College London, London SW7 2BW, UK \\
         $^{2}$ The Cavendish Laboratory, Madingley Rd, Cambridge CB3 0HE, UK \\
         $^{3}$ School of Physics and Astronomy, University of Nottingham,
         NG7 2RD, UK
}

\recdate{\today}

\abst
{
A which-way device is one which is designed to detect which of 2 paths
is taken by a quantum particle, whether Schr\"odinger's cat is dead or
alive.  One possible such device is represented by
an Aharonov--Bohm ring with a quantum dot on one branch.  A charged cantilever or
spring is brought close to the dot as a detector of the presence of an
electron.  The conventional view of such a device is that any change in
the state of the cantilever implies a change in the electron state which
will in turn destroy the interference effects.
In this paper we show that it is in fact possible
to change the state of the oscillator while preserving the quantum interference
phenomenon.
}

\kword{%
NEMS; mesoscopics; Aharonov--Bohm; which--way
device}

\begin{document}
\sloppy
\maketitle

\newcommand{\wee}[1]{{\mbox{\scriptsize\rm #1}}}
\newcommand{\weee}[1]{{\mbox{\tiny\rm #1}}}
\newcommand{\xe}{x_\wee{e}}
\newcommand{\xee}{x_\weee{e}}
\newcommand{\yc}{y_\wee{c}}
\newcommand{\kc}{k_\wee{c}}

\section{Introduction}
The concept of a ``which--way'' device has always played an
important role in our understanding of quantum mechanics,
representing as it does one of the most difficult concepts in
modern physics: Schr{\"o}dinger's famous cat is neither alive nor
dead.  Textbooks of quantum mechanics typically discuss electrons
going through a double slit and state that any attempt to identify
which slit the electron goes through will result in the
destruction of the interference pattern associated with the double
slit\cite{rpf}.

In a recent experiment\cite{Buks98}, Buks {\it et al.} used a
``which-way'' device to probe dephasing effects in mesoscopic
electronic systems. They fabricated  an Aharonov--Bohm
ring\cite{AhaBohm59,Webb+85} in a semiconductor structure with a
quantum dot included in one of the arms. A quantum point contact
(QPC) was fabricated adjacent to the dot so that the current
flowing through it was modulated when the nearby dot was occupied.
In this system the current through the QPC effectively measures
the path taken by the electron (by measuring the occupancy of the
adjacent quantum dot). Measurement of the path taken by an
electron in an Aharonov--Bohm ring inevitably leads to dephasing
of the electrons, and hence suppression of the interference
fringes in the current as a function of flux.

\begin{figure}[ht]
\begin{center}\leavevmode
\includegraphics[width=0.6\linewidth]{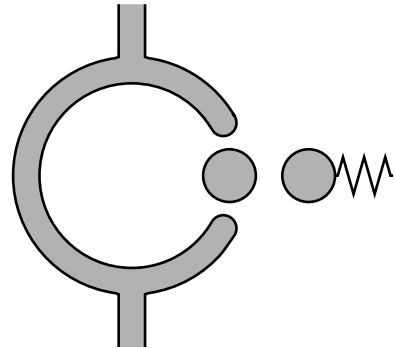}
\caption{
An Aharonov--Bohm ring containing a quantum dot in close proximity to another
charged dot attached to a spring or cantilever.
}\label{fig:whichway}\end{center}
\end{figure}
Recently, Armour and Blencowe\cite{ArmBlen01,ArmBlen02} discussed
the electromechanical ``which-path'' device illustrated in
figure~\ref{fig:whichway}. This device is based on the system
investigated by Buks {\it et al.}, but with the QPC replaced by a
micron-sized cantilever, positioned close to the dot. The
cantilever can be coated in a thin metal layer so that it carries
a net charge and therefore interacts with charges on the
dot\cite{cantdot}. The presence of an additional electron on the
dot couples effectively to just the fundamental flexural mode of
the charged cantilever\cite{ArmBlen01}, causing a deflection. For
low enough temperatures, and for high enough resonant frequencies,
the fundamental mode of the cantilever can be treated as a quantum
harmonic oscillator and coherent coupling between the electrons
and the cantilever must be considered\cite{ArmBlen01}.

Considering a cantilever coupled to the dot rather than a QPC is
of interest for two main reasons. Firstly, on a practical level,
coupling to an Aharonov--Bohm ring may provide a very effective
way of probing quantum behaviour in micro-mechanical
systems\cite{ArmBlenS}. Secondly, because the problem reduces to
that of coupling to a quantum harmonic oscillator, it provides a
generic model for investigating systems in which the electrons
couple to a single, localised degree of freedom (in contrast to
the many, transient, degrees of freedom associated with electrons
passing through a QPC).

The conventional description of what happens in the
electromechanical ``which--path'' device would be that the
electrons are dephased whenever their interaction with the
cantilever results in a change in the state of the cantilever.
However, this presupposes that the electrons travel no more than
half-way around the ring before leaving at the top junction so
that only one of the two interfering paths passes through the dot.
In practice, for a device in which the only connections between
the ring and the outside world occur via the top and bottom
leads\cite{note}, many electron paths contribute to the current
through the device, some of which travel around the ring several
times (therefore passing through the dot several
times)\cite{yacoby96}.

The presence of interfering paths which each pass through the
quantum dot leads to interesting effects which arise from the fact
that the electrons can couple to the cantilever coherently.
Because the electrons on the dot interact effectively with just
the fundamental flexural mode of the cantilever\cite{ArmBlen01},
it is possible for electrons passing through the dot on different
paths to change the state of the cantilever {\it in the same way}
and so remain phase coherent.

The idea that electrons can undergo inelastic scattering and still
remain phase coherent goes against the intuitive picture developed
in many text-books that a change of state in the measuring device
or environment always causes dephasing. However, it has been
recognised for some time\cite{Imry} that exchange of energy is
neither a necessary nor a sufficient condition for dephasing.

In this paper we develop a numerical simulation to investigate the
nature of the interference fringes in the current as a function of
the magnetic flux when the electron paths can thread the ring more
than once. In particular we explore the contribution to the
fringes arising from interfering electron paths which include
inelastic scattering.

\section{The Model}
The system may be described by a Hamiltonian of the form
\begin{eqnarray}
H &=&
-E_{\rm e}\left({\partial\over\partial \xe}
- e{\Phi\over 2\pi R}\right)^2\nonumber\\
&&- E_{\rm c}\left({\partial^2\over\partial \yc^2} -
{\textstyle\frac14}\yc^2\right)+ V(\xe, \yc)\label{eq:hamil}
\end{eqnarray}
where $\xe$ and $\yc$ are the electron and phonon
(i.e.\ the fundamental flexural mode of the cantilever)
coordinates respectively, $\Phi$ and $R$ are the magnetic flux and
the radius of the ring and
\newcommand{\negspace}{\hspace{-10pt}}
\begin{equation}
V(\xe, \yc) = \left\{\begin{array}{crl}
                           \alpha \yc & &|\xe| < a\\
                          V_{\rm b} & a<\negspace&|\xe|< a+d\\
                            0 & a+d <\negspace&|\xe|
                          \end{array}\right. \;,
\end{equation}
which contains a pair of barriers of height $V_{\rm b}$ and width $d$ separated by
$2a$.  The depth of the well is $\alpha \yc$ which represents the
coupling to the cantilever.  Thus the effective potential inside the well
may be written as
\begin{equation}
V_{\mbox{\scriptsize\rm eff}}
= \frac{E_{\rm c}}4\left[\left(y_c+\frac{2\alpha}{E_{\rm c}}\right)^2
-\left(\frac{2\alpha}{E_{\rm c}}\right)^2\right]
\end{equation}
which is a harmonic potential shifted with respect to that outside the
well.
The matching of the wave functions at the boundaries of the well is most
easily carried out by Fourier transforming with respect to $\yc$ so that
the $n$--phonon state inside of the well $\phi_n^\wee{well}(\kc)$ is related
to that outside by
\begin{equation}
\phi_n^\wee{well}
= \exp\left({\rm i}\kc\frac{2\alpha}{E_\wee{c}}\right)\phi_n^\wee{out}\;.
\label{eq:kshift}
\end{equation}
where the exponential can be turned into an operator in the phonon number
representation by first noting that we can write
\begin{equation}
\hat k_c = \textstyle{\frac12}{\rm i}\left(\hat c^\dagger - \hat c\right)
\end{equation}
where $\hat c^\dagger, \hat c$ are the phonon creation and annihilation operators.
This operator may be written in matrix form and diagonalised, the exponential
carried out on the eigenvalues, and the result transformed back into the
phonon number representation. It is only necessary to perform this operation
once, at the start of the calculation.

Writing the wave functions in the form
\begin{equation}
\psi(\xe,\kc) = \sum_n \phi_n(\kc)\left( a_n^+ e^{{\rm i} k_n^+ \xee} +
a_n^- e^{{\rm i} k_n^- \xee}\right)
\end{equation}
where $k_n^\pm$ are the left and right--going wave vectors of the electronic
part of the wave function when the cantilever is in the $n$--phonon state
(in the absence of the magnetic flux $k_n^- = -k_n^+$)
it is straightforward to derive expressions for the transmission and reflection
matrices at various boundaries in the potential and to combine these into
transmission and reflection coefficients, and hence scattering ($S$) matrices
describing the whole system.

At the junctions with the leads amplitude and
current conservation are ensured by appropriate boundary
conditions
\begin{eqnarray}
&&\psi_1 = \psi_2 =\psi_3\nonumber \\
&&\left.d\psi\over dx\right|_1 + \left.d\psi\over dx\right|_2 + \left.d\psi\over dx\right|_3
= 0 \nonumber
\end{eqnarray}
where the derivatives are all defined towards the junction.
In terms of scattering matrices $\mathbf{S}$ which describe the coupling of
incident and emitted waves we can combine the 2 branches into an Aharonov--Bohm
loop using
\begin{equation}
\mathbf{\Sigma}_{12}=\mathbf{\Sigma}_1 + \mathbf{\Sigma}_{2}
\end{equation}
where
\begin{equation}
\mathbf{\Sigma} = (\mathbf{S} -1)*(\mathbf{S} +1)^{-1}
\end{equation}

\section{Results}
\begin{figure}[ht]
\begin{center}\leavevmode
\includegraphics[width=\linewidth]{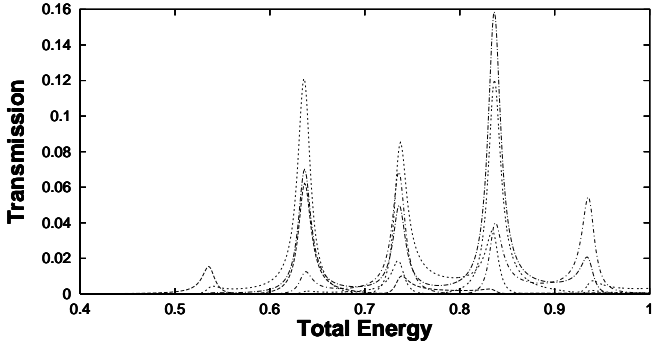}
\caption{ Transmission coefficients of the dot as a function of
total energy ($\mbox{electron} + \mbox{phonon}$) with the
cantilever in initial state $n=2$.  The various lines represent
the final state of the cantilever  ($n=0\quad\mbox{-- -- --}$, 
$n=1\quad\mbox{- - -}$,$n=2\quad\mbox{--}\cdot\mbox{--}\cdot$, 
$n=3\quad\mbox{-}\cdot\mbox{-}\cdot$, $n=4\quad\cdot\cdot\quad\cdot\cdot$).
The parameters used are: $E_{\rm e}=1$, $E_\wee{ph}=0.1$,
$\alpha=0.05$, $V_b=4$, $d=1$, $2a=\pi$. }\label{fig:cantoff2}
\end{center}
\end{figure}
For orientation fig.~\ref{fig:cantoff2} shows the transmission of
the dot without the ring as a function of total energy for the
parameters used throughout this paper. In particular the
cantilever is initially in the 2 phonon state ($n=2$) but the
different curves correspond to different final states of the
cantilever. Note the well defined resonant tunnelling peaks
separated by the phonon energy $E_\wee{ph}$ corresponding to the
bound state of the dot with $n$ phonons.  Note also the strong
mixing of the cantilever states illustrated by the presence of
finite weight in all peaks for various final states.  We have, of
course, chosen a relatively strong coupling for the purpose of
illustration, but not so strong as to introduce new physics.

\begin{figure}[ht]
\begin{center}\leavevmode
\includegraphics[width=\linewidth]{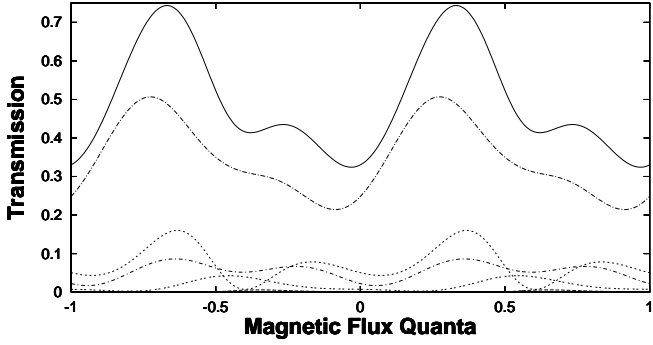}
\caption{ Transmission coefficients of the the ring as a function
of magnetic flux with the cantilever in an initial state with 2
phonons.  The energy is $0.7365$, the maximum of the 3rd peak in
fig.~\ref{fig:cantoff2}\cite{onsagernote}. 
The labelling and parameters are as in
fig.~\ref{fig:cantoff2}.  

}\label{fig:peak2off2}
\end{center}
\end{figure}
Fig.~\ref{fig:peak2off2} shows the transmission of the ring as a
function of magnetic flux (in units of $h/e$) again with the
cantilever initially in the 2 phonon state. The energy is chosen
as $0.7365$ which is the maximum of the 3rd peak in
fig.~\ref{fig:cantoff2} corresponding to the bound state of the
dot plus 2 phonons.  Note that the $n=2$ line, which corresponds
to the elastic transmission in which the state of the cantilever
is unchanged, has higher values than the other curves but is still
modulated by the magnetic field.  In the
absence of the cantilever and the ring the transmission here would
be perfect.  It is reduced from this value due to scattering by
the cantilever and interference effects on the ring. This curve
clearly contains components with period of $1$ corresponding to
the usual Aharonov--Bohm effect\cite{AhaBohm59} and usually
interpreted as interference between the 2 different paths joining
the source and the drain leads of the ring. The curve is far from
perfectly sinusoidal however, indicating the presence of higher
order contributions such as the weak localisation or correlated
back scattering effect \cite{macK01} corresponding to interference
between paths round the whole ring in clockwise and anticlockwise
directions respectively.

The curves in fig.~\ref{fig:peak2off2} with different final states, corresponding to the
creation or annihilation of one or more phonons have significant weight
and, perhaps unexpectedly, are also modulated by the magnetic field.  These
results correspond to a situation in which the state of the cantilever
has been changed but the transmission still shows interference effects.
Is this consistent with the idea that a change of state of a measuring
device causes dephasing and hence destroys interference? The na{\"\i}ve
interpretation would be that the cantilever's change of state indicates
that the electron must have gone through the dot and emerged with a different
energy.  This would not then interfere with a wave which has gone by the alternative path and
has not gone through the dot.

\begin{figure}[ht]
\begin{center}\leavevmode
\includegraphics[width=\linewidth]{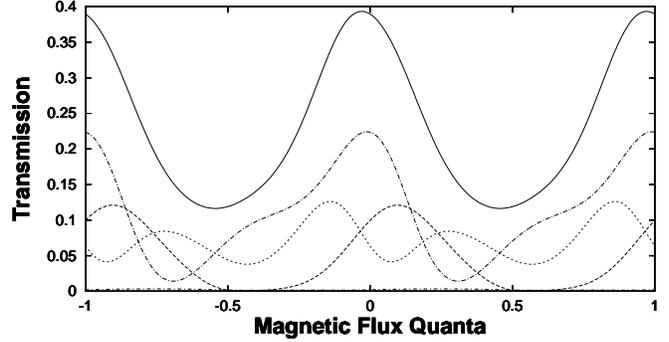}
\caption{ Transmission coefficients of the the ring as a function
of magnetic flux (in units of $h/e$) with the cantilever in an
initial state with 2 phonons.  The energy is $0.6365$, the maximum
of the 2nd peak in fig.~\ref{fig:cantoff2}\cite{onsagernote}. The labelling and
parameters are as in fig.~\ref{fig:cantoff2}.
}\label{fig:peak1off2}
\end{center}
\end{figure}
In order to investigate this further we change the energy to correspond
to the 2nd peak in fig.~\ref{fig:cantoff2} at the bound state of the
dot but with a single phonon.  The initial state still has 2 phonons.  Now we
observe that there are several curves with similar average transmission
probabilities and similar amplitudes of modulation. In fact no particular
state is dominant.  This is not so surprising
given that the energy chosen corresponds to a different state of the dot than
would correspond to the initial state of the cantilever.
Both the $n=0$ and $n=2$
curves are dominated by contributions with periods of 1 flux quantum $h/e$, but
the $n=1$ curve clearly has a significant contribution with $\textstyle{\frac12}h/e$.

\begin{figure}[ht]
\begin{center}\leavevmode
\includegraphics[width=\linewidth]{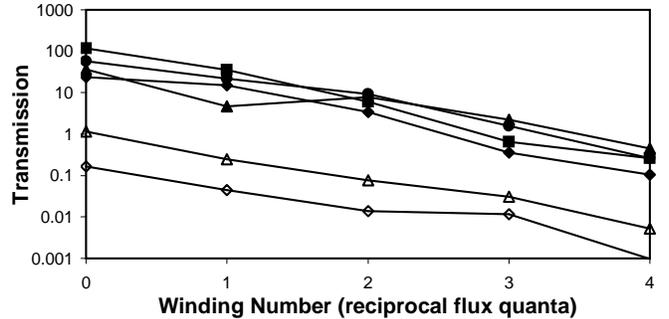}
\caption{The Fourier transform of fig.\ref{fig:peak1off2} plotted on a
logarithmic scale for clarity.  Note the presence of structure at various
periods (units of $e/h$).  Initial state $n=2$; total transmission - squares,
final state $n=0$ - filled diamond, $n=1$ - filled triangle,
$n=2$ - filled circle, $n=3$ - open triangle, $n=4$ - open diamond. Note that the
lines are a guide to the eye and should not be interpreted as intermediate
values. As the calculation
is performed with perfectly 1D wires the results are perfectly periodic
and the Fourier transform is exactly zero between the plotted values.
}\label{fig:fourieroff2}
\end{center}
\end{figure}
This becomes clearer when the data in fig.~\ref{fig:peak1off2} are
Fourier transformed. In fig.~\ref{fig:fourieroff2} the $n=1$
results (filled diamonds) have larger higher order components, but there
are also significant contributions at other phonon numbers. Again
we observe quantum interference in spite of the change of state of
the cantilever.

\section{Analysis}
\begin{figure}[ht]
\begin{center}\leavevmode
\includegraphics[height=0.4\linewidth]{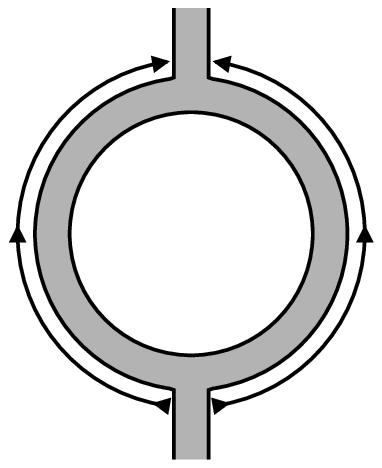}\hfil
\includegraphics[height=0.4\linewidth]{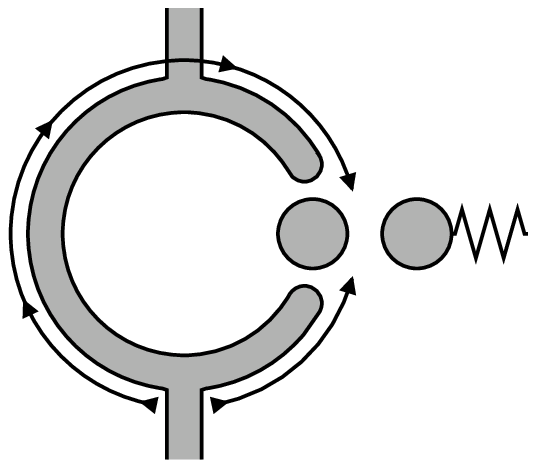}
\caption{
Left: a conventional Aharonov--Bohm ring, with an indication of the 2 partial
waves which interfere to give a conductance oscillation with period $h/e$.
Right: the device discussed here, with an indication of the 2 partial waves
which interfere at the quantum dot.
}\label{fig:whichway2}\end{center}
\end{figure}
How is it possible for us to observe interference at the same time
as a change of state of the cantilever? The first important
observation is that the period of the most significant
contribution is a single flux quantum $h/e$, similar to the
original Aharonov--Bohm effect\cite{AhaBohm59}. The ``winding
number'' of the responsible interference process must therefore be
one. The simplest such process involves interference between partial waves
going clockwise and anti--clockwise {\it at the dot\/}, rather
than at the source or drain of the ring (see figure~\ref{fig:whichway2}.   
Thus the amplitude of
the wave function at the dot varies with a period of 1 flux
quantum. The value of the matrix element coupling the electron to
the cantilever is proportional to this amplitude such that the
strength of the coupling between the electron and the cantilever
also varies with period $h/e$.  As this accounts for the basic
periodicity of the process the waves emitted from the dot are not
subject to any further interference and presumably may
be described by a single wave travelling from the dot to the drain
of the ring.

On the other hand a process might be considered by which a single wave
travels from the source to the dot where a phonon is created or annihilated.
The emitted wave may then split in 2 and travel either clockwise or anti--clockwise
to the source or to the drain, where the 2 partial waves again interfere.  Note
that this process also has a winding number of unity.

A simple higher order process involves a combination of the above 2 paths;
2 paths from the source to the dot and 2 paths from the dot to the source
or drain. In such a process the first step would have periodicity
$h/e$ but the source of the 2nd step is itself modulated with the same periodicity. If
the basic process is controlled by a term $\propto\cos\left(2\pi \frac{e}{h}\Phi\right)$
then the 2nd order one has an amplitude
$\propto\cos^2\left(2\pi\frac{e}{h}\Phi\right)$ which has period $h/2e$.

It would be interesting to be able to filter out the elastic contribution.
For the particular configuration discussed here a simple method for doing so would be 
to add an extra barrier to the
drain lead with a height between the $n=0$ and $n=1$ bound state energies
of the dot. This would block the electron if it retains its initial energy
but allow it through if it has gained $E_\wee{c}$ by annihilation of a
phonon.  More generally the behaviour of the system as a function of the
height of such a barrier would be a useful probe of the effect.

\section{Discussion}
Why do these inelastic processes not destroy the quantum
interference? In most discussions of dissipation the energy
transfer is with a continuum such that the final state of the
electron is also a continuum\cite{CaldLegg83}. In the present case
the coupling is to a single vibrational mode and the final state
is therefore a linear combination of well defined separate states.
Thus interference is still possible even after the energy has been
changed. Of course the processes which destroy the conventional
Aharonov--Bohm effects will still be operative here\cite{hansen01}
and will tend to suppress higher order processes involving
multiple circuits of the ring.

The fact that the transmission can be dominated by contributions
involving a change in phonon number with a periodicity of one flux
quantum means that the distance the electrons have to travel
through the ring coherently is the same as for the usual elastic
processes. This  implies that the lowest-order inelastic
interference effects should be no more difficult to observe than
the elastic ones, discussed in detail by Armour and
Blencowe\cite{ArmBlen01}. The key practical difficulty is in
obtaining reasonably strong coupling between the quantum dot and a
cantilever of sufficiently high frequency that the thermal
broadening of the electron energies in the leads does not exceed
the energy of the phonons associated with the fundamental flexural
mode.

\begin{figure}[ht]
\begin{center}\leavevmode
\includegraphics[width=\linewidth]{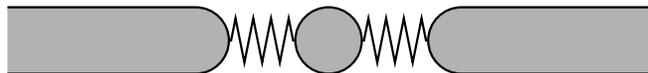}
\caption{An example of a quantum shuttle in which a quantum dot
can oscillate between 2 contacts.}\label{fig:shuttle}
\end{center}
\end{figure}
An interesting modification of the system we have considered here
(shown in fig.~\ref{fig:whichway}) would be to combine the quantum
oscillator and the dot into one, forming a quantum
shuttle\cite{mKArm02} as in fig.~\ref{fig:shuttle}. Although the
essential geometry of the system would remain the same, this
alternative set-up might well make it easier to couple the
electronic degrees of freedom to a very high frequency
oscillator\cite{possible}.

We have shown that the which-way device involving coupling to a nanoscopic
cantilever, which is a simple modification
of an Aharonov--Bohm ring, has unexpected
properties.  It is possible to change the state of the detector, the cantilever,
while retaining the quantum interference.  This appears to contradict
the statement contained in many textbooks that any attempt to determine
which path the electron has taken will result in the destruction of the
interference. However, the nature of the process
involved is such that it actually tells us absolutely nothing about which path the
electron has taken.  In fact it confirms that the electron has taken both
paths.
Hence the conventional view is not wrong; it just requires a more subtle interpretation.

\section*{Acknowledgements}
We would like to thank the EPSRC for financial support and the Cavendish Laboratory,
Cambridge for its hospitality.


\begin{thebibliography}{99}
\bibitem{note} Note that this is not the case in the experiment of Buks {\it et al.}
and the device analysed by Armour and Blencowe.
\bibitem{onsagernote} Note that the quantity plotted here
is $\left|t_{ij}\right|^2$ for electrons incident from one lead.  This
is equal to the corresponding quantity for electrons incident from the
other lead only when the magnetic field is reversed.  The conductivity
contains the sum of both quantities and is thus symmetric in the field
(Onsager relation).
\end{thebibliography}
\end{document}